\documentclass[aps,12pt,notitlepage,final,oneside,onecolumn,nobibnotes,
nofootinbib,superscriptaddress,showpacs,centertags,amsmath,amssymb]{revtex4}

\begin{document}

\title{Fermion resonance in quantum field theory}
\author{\firstname{M.O.} \surname{Gonchar}}
\author{\firstname{A.E.} \surname{Kaloshin}}
\email{kaloshin@physdep.isu.ru}
\author{\firstname{V.P.} \surname{Lomov}}
\email{lomov@physdep.isu.ru}

\affiliation{Russia, Irkutsk State University}

\begin{abstract}
  We derive accurately the fermion resonance propagator by means of Dyson
  summation of the self-energy contribution. It turns out that the
  relativistic fermion resonance differs essentially from its boson analog.
\end{abstract}


\maketitle

\section{Introduction}

The Breit-Wigner formula describing the processes of production
and decay of unstable particle is widely used in hadron and
nuclear physics. The original formula \cite{BW}, which was applied
to scattering of slow neutrons, is a non-relativistic one with
energy-independent parameters -- resonance position and width. It
is clear that the simplest Breit-Wigner formula is applicable only
for narrow states and description of a resonance curve for broad
hadron resonances (at high experimental accuracy) needs some
improved resonance formula. In hadron phenomenology there exist
slightly different methods to describe resonance contribution but
practically all methods are based on (effective) quantum field
theory.

In spite of long history  we found with some astonishment that
situation with description of fermion resonance is not so evident
up to now. Naively one can expect that the fermion resonance
contribution to cross-section is the same as the boson one. Such
point of view may be found in many papers and
reviews, in particular, in \cite{Yao}%
\footnote{There is no common opinion concerning the form of the
relativistic fermion propagator for resonance. But all variants found by us, e.g.
\cite{Pas,Lop,Alv,Ani}, are in fact some simplification of general
formula \eqref{BWF}. As for resonance denominator, it is always
written in analogy with boson case \eqref{BW} but not in correct
form \eqref{BWF}. Of course, there is a possibility to avoid the
problem: to use the $W$-dependent partial waves and to forget
about symmetry properties $W\to -W$. Such way is used, in
particular, in $\pi N$  partial analysis \cite{Hoh,Cut,Arn04}.} .%
 But this is not the case, as it is seen from our consideration.
First of all, the natural variable  in fermion case is
$W=\sqrt{s}$ but not $s$, so the fermion resonance factor will be
$W$-dependent. After renormalization of dressed propagator one can see that the fermion 
specifics is generated by antisymmetric in $W$ contributions in a self-energy. This property
is easily seen with use of the off-shell projection operators instead of $\gamma$-matrices.

Below we derive the Breit-Wigner-like formula for fermions in
quantum field theory and discuss some its properties. We found
that the so obtained dressed propagator \eqref{BWF} has some
interesting features specific for fermions.

\section{Boson resonance in QFT}

Let us first consider  a more evident case of the boson resonance.
The unstable particle is usually associated with the Breit-Wigner
formula for an amplitude
\begin{equation}\label{BW}
  T(a\to R\to b)= g_a \cdot \dfrac{1}{M_R^2 -s -i\Gamma_R M_R} \cdot g_b .
\end{equation}
Here the factor $G(s)=1/(M_R^2 -s -i\Gamma_R M_R)$ represents relativistic propagator of
unstable particle.

The similar formula may be obtained in framework of quantum field
theory by means of Dyson summation of the self-energy insertions
into propagator. Equivalently, we should solve the Dyson-Schwinger
equation for dressed nonrenormalized propagator
\begin{equation}\label{DS}
  G=G_0 + G J G_0 .
\end{equation}
Here $G_0(s)=(M^2 -s)^{-1} $ and $G(s)$ are the bare and total
propagators respectively, $J(s)$ is the self-energy contribution
(sum of the 1PI diagrams).

The equation \eqref{DS} may be rewritten in terms of inverse
propagators $S(s)=G^{-1}(s)$%
 \footnote{If we consider the
self-energy $J(s)$ as a known value (i.e.we neglect by dressing of
  the vertex), then we have so called rainbow approximation, see e.g.
  review \cite{Mar}. This
  approximation is sufficient to respect the analytical properties
  and unitarity and is widely used in resonance physics. Sometimes
  this approach is not sufficient and people use some
  post-rainbow approximations. As an example we can remember the
  well known trick with introducing
  of the centrifugal barrier factor into width.}
\begin{equation}\label{DS1}
  S = S_0 - J = M^2 -s - J(s) .
\end{equation}
If to use the on-mass-shell scheme of renormalization, then $M$ is
the renormalized mass and one should subtract  the self-energy
contribution twice at this point
\begin{equation}\label{DS2}
  S^{r} = M^2 -s - [ J(s)- \Re J(M^2) - \Re J^{\prime}(M^2)(s-M^2) ].
\end{equation}
After it we have formula similar to \eqref{BW} but with "running"
mass and width. From QFT point of view the Breit-Wigner formula is
rather rough approach, when we neglect by the energy dependence in
mass and width (i.e. by the energy dependence in loop contribution
$J(s)$). Such simplification is justified only for narrow
resonance located far from threshold.

\section{Fermion resonance in QFT}

\subsection{Dressing and renormalization}

Dressing of the fermion propagator looks rather similar:
\begin{equation}\label{R12}
  G_0(p)=\frac{1}{\hat{p}-M}\ \  \Rightarrow \ \
  G(p)=\frac{1}{\hat{p}-M -\Sigma(p)},
\end{equation}
where $\Sigma(p)=A(p^2)+\hat{p} B(p^2)$ is a self-energy contribution.

To renormalize the obtained expression it is convenient to work
with the inverse propagator $S(p)=G^{-1}(p)$.  Usually the
renormalization condition is formulated as the decomposition of
inverse propagator in terms of $\hat{p}-M$
\begin{equation}
  S(p) = \hat{p} - M + o(\hat{p} - M)\quad \text{at}\quad\hat{p}\to M .
\end{equation}

Useful technical step is to introduce the basis of the off-shell projection operators
$\Lambda^{\pm}$
\begin{equation}\label{L}
  \Lambda^{\pm}=\dfrac{1}{2}\big( 1 \pm \frac{\hat{p}}{W} \big),\quad W=\sqrt{p^2},
\end{equation}
which simplifies all operations with $\gamma$-matrices and makes
more clear the renormalization procedure. For illustration, let us
use this basis for the self-energy contribution. Evident chain
gives:
\begin{eqnarray}
  \Sigma(p)&=&A(p^2)+\hat{p}B(p^2)=(\Lambda^{+}+\Lambda^{-})A(W^2) +
  W(\Lambda^{+}-\Lambda^{-}) B(W^2)\nonumber \\
  &=& \Lambda^{+} [A(W^2)+W B(W^2)] + \Lambda^{-} [A(W^2)-W B(W^2)].
\end{eqnarray}
One can see that coefficients $\Sigma^{\pm}$ in this basis are related by
simple substitution
$\Sigma^-(W)=\Sigma^+(-W)$ and dressed propagator has this property also.

The reversing of propagator is very easy due to simple multiplicative properties of the
basis. If the inverse propagator has a decomposition
\begin{equation}
  S(p)=\Lambda^{+} S^+(W)+\Lambda^{-} S^-(W) ,
\end{equation}
with the symmetry property $S^-(W)=S^+(-W)$, then the propagator $G(p)$ is of the form
\begin{equation}\label{}
  G(p)=\Lambda^{+}\frac{1}{S^+(W)}+\Lambda^{-}\frac{1}{S^-(W)}.
\end{equation}
So, the explicit form of dressed unrenormalized propagator is evident:
\begin{equation}\label{}
  G(p)=\Lambda^{+}\dfrac{1}{W-M-(A(W^2)+W B(W^2))}+\Lambda^{-}\dfrac{1}{-W-M-(A(W^2)-W B(W^2))}.
\end{equation}
Thus, using the $\Lambda$-basis we have separated the
$\gamma$-matrix structure and should renormalize the scalar
coefficients dependent on $W$. More exactly, we need renormalize
only $G^+(W)$ component, after that another coefficient $G^-(W)$
is obtained by substitution $W\to -W$.

For bound state, located below the threshold, the
renormalization leads to the following condition for the
self-energy contribution:
\begin{equation*}
  S^{+}=W-M+o(W-M) \quad\text{at}\quad W \to M  .
\end{equation*}
One can convince yourself that the final expression, obtained by
using  the projection operator basis, coincides with the
standard one presented in any textbook.

If we deal with a resonance located higher the threshold, the
renormalization condition takes the form:
\begin{equation}
  S^{+}=W-M+o(W-M)+i\frac{\Gamma}{2} \quad\text{at}\quad W \sim M .
\label{renorm}
\end{equation}
Note that the real part of \eqref{renorm} is some requirement on the
subtraction constants of self-energy functions $A(p^2), B(p^2)$:
\begin{equation}
  \label{a3}
  \begin{split}
    \Re A(M^2)+M\Re B(M^2)=0, \\
    2M \Re A^{\prime}(M^2)+\Re B(M^2)+2M^2 \Re B^{\prime}(M^2)=0 ,
  \end{split}
\end{equation}
whereas the imaginary part of the condition \eqref{renorm} simply
relates the coupling constant and width%
 \footnote{It is known that
normalization on the pole in complex energy plane is preferable
from theoretical point of view \cite{Sir,Pass} but for our purpose
the more crude recipe
  \eqref{renorm} is sufficient.}.
\begin{equation}
  -(\Im A(M^2)+M \Im B(M^2)) = \frac{\Gamma}{2}.
\end{equation}
Eq. \eqref{a3} fixes the loop subtraction constants and after it
the functions $A(W^2), B(W^2)$ are defined completely. Inverse
propagator may be written in the form similar to Breit-Wigner formula
\eqref{BW} but with "running" mass and width
\begin{equation}
  S^{+}(W) = W-M(W)+\frac{i}{2}\Gamma(W) . \label{BW1}
\end{equation}
Another component is obtained by the substitution $W\to -W$:
\begin{equation}
  S^{-}(W) = -W-M(-W)+\frac{i}{2}\Gamma(-W) . \label{BW2}
\end{equation}
If to look at the self-energy contribution, one can see that there
are the symmetric and antisymmetric contributions under the $W\to
-W$ exchange. Therefore the running mass and width also may be
divided into two parts:
\begin{equation}
  M(W)=M^S(W)+M^A(W),\qquad \Gamma(W)=\Gamma^S(W)+\Gamma^A(W)
\end{equation}
and components of renormalized propagator take the form:
\begin{equation}
  \begin{split}
    S^{+}&=W-M^S(W)-M^A(W)+\frac{i}{2}(\Gamma^S(W)+\Gamma^A(W)), \\
    S^{-}&=-W-M^S(W)+M^A(W)+\frac{i}{2}(\Gamma^S(W)-\Gamma^A(W)).
  \end{split}
\end{equation}
Let us stress that these components are normalized at different points:
\begin{equation}\label{normal}
  \begin{split}
    S^+ &\approx W - M + \frac{i}{2} \Gamma \quad\quad \ \ \mbox{at}\ W \sim M \\
    S^- &\approx -W - M + \frac{i}{2} \Gamma \quad\quad \mbox{at}\ W \sim -M .
  \end{split}
\end{equation}

Returning from projection operators to $\hat{p},I$ basis, we
obtain the following formula for the resonance propagator:
\begin{equation}\label{BWF}
  G(p)=\dfrac{1}{\Delta}\left[ M^S(W)-\frac{i}{2}\Gamma^S(W)+\hat{p}\left(1-\frac{M^A(W)-
        \dfrac{i}{2}\Gamma^A(W)}{W} \right)\right],
\end{equation}
where
\begin{equation*}
  \Delta(W^2)=\left( W-M^A(W)+\dfrac{i}{2}\Gamma^A(W) \right)^2-\left( M^S(W)-\frac{i}{2}\Gamma^S(W) \right)^2.
\end{equation*}

Let us compare it with boson Breit-Wigner formula for the inverse
propagator $S(p^2)$ written in a similar form:
\begin{equation}
  S(s)=s-M^2(s)+iM\Gamma(s)\approx W^2-(M-i\frac{\Gamma}{2})^2=
       [W-(M-\frac{i}{2}\Gamma)][W+(M-\frac{i}{2}\Gamma)].
\end{equation}
One can see that the fermion denominator in \eqref{BWF} turns into
boson one in absence of antisymmetric contributions:\
$M^A=\Gamma^A=0$.

Let us compare also the interference pictures near the positive energy pole in elastic
amplitude.

\noindent\underline{\textbf{Boson amplitude}}
\begin{equation}\label{intb}
  \mathcal{T}=\dfrac{g^2}{M^2 -s -iM\Gamma}\approx\frac{-g^2}{2M}\left(\frac{1}{W-M+i\Gamma/2}
              +\frac{1}{-W-M + i\Gamma/2} \right)
\end{equation}

\noindent\underline{\textbf{Fermion amplitude}}

For definiteness let us consider the process $\pi N\to
N^{\prime}(1/2^+) \to \pi N$ and look at the center mass helicity
amplitude $\mathcal{M}_{++}$ at $W\sim M$
\begin{equation}\label{intf}
  \mathcal{M}_{++}= \cos{\frac{\theta}{2}}\ \left[\dfrac{(p^0+m_N)g^2}{-W-(M^S - M^A)+
                   i(\Gamma^S-\Gamma^A)/2} - \dfrac{(p^0-m_N)g^2}{W-M+i\Gamma/2}\right],
\end{equation}
where $p^0$ is the nucleon c.m.s. energy.

One can see that in contrast to boson case the background
contribution in vicinity of $W=M$ is not expressed in terms of $M$
and $\Gamma$. This feature is generated by $\hat{p}$ contribution in
self-energy $\Sigma(p)=A(p^2)+\hat{p} B(p^2)$. As a result instead
of two parameters ($M$ and $\Gamma$) the fermion Breit-Wigner is
described by four parameters (one can use $M$,$\Gamma$ and complex
background).

All above formulae are applicable directly to spin-3/2 resonances.
The dressed propagator of Rarita-Schwinger field has the form
\cite{KL,KL1} (compare with \eqref{R12}):
\begin{equation}\label{R32}
  G^{\mu\nu}(p)=\dfrac{1}{\hat{p}-M -\Sigma(p)}\mathcal{P}_{3/2}^{\mu\nu} + (\text{spin-1/2 terms}),
\end{equation}
where the operator $\mathcal{P}_{3/2}^{\mu\nu}$ can be found in
\cite{Nie}. One can see that all above operations are repeated
with the $1/(\hat{p}-M -\Sigma(p))$ factor, which after all takes
the form \eqref{BWF}.

\begin{figure}
  \centering
  \includegraphics*[width=8cm]{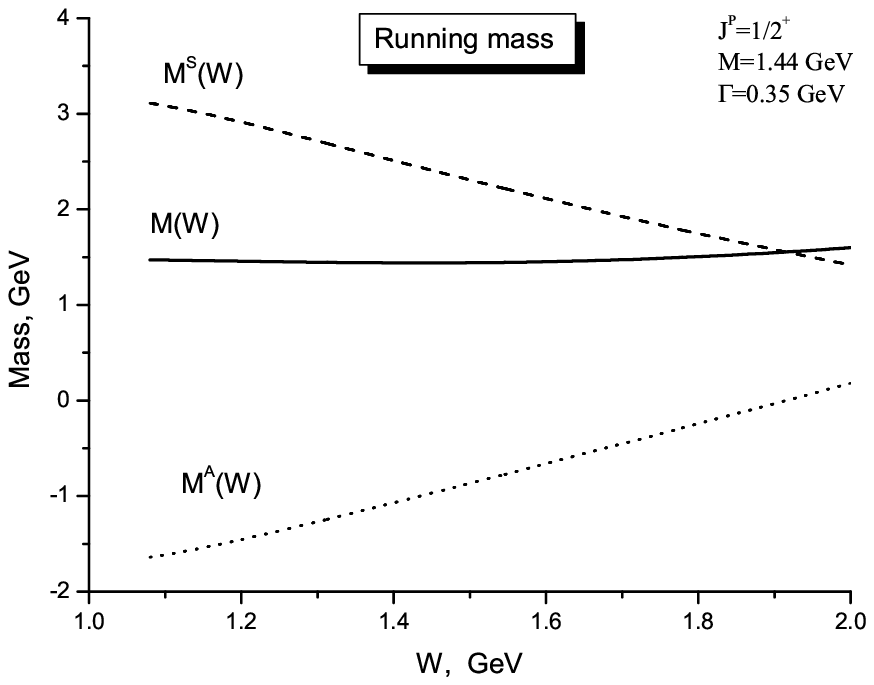}
  \includegraphics*[width=8cm]{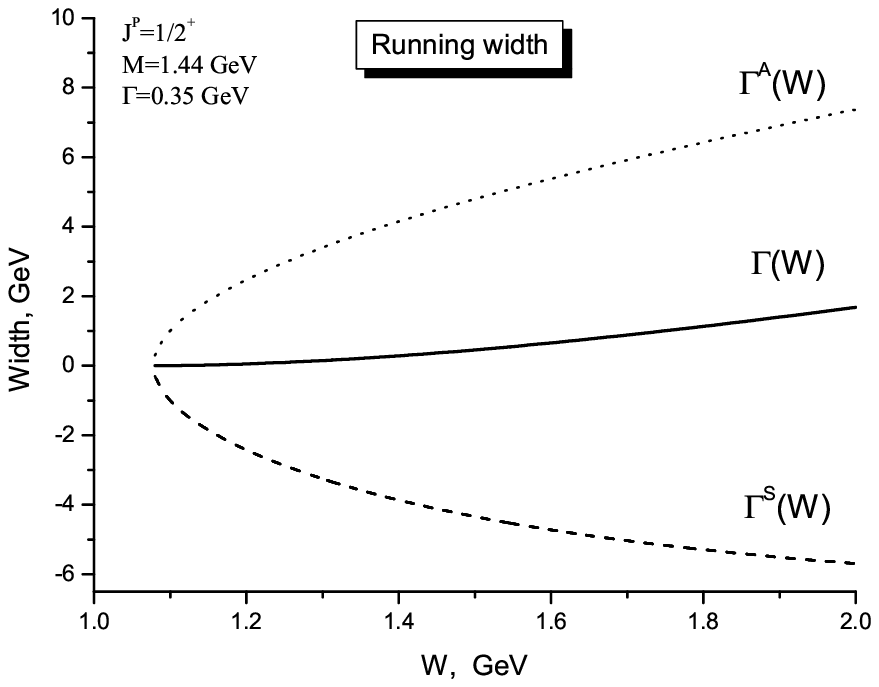}
  \caption{\label{fig1}Running mass and width and their components for resonance of
positive parity.}
\end{figure}

\begin{figure}
  \centering
  \includegraphics*[width=8cm]{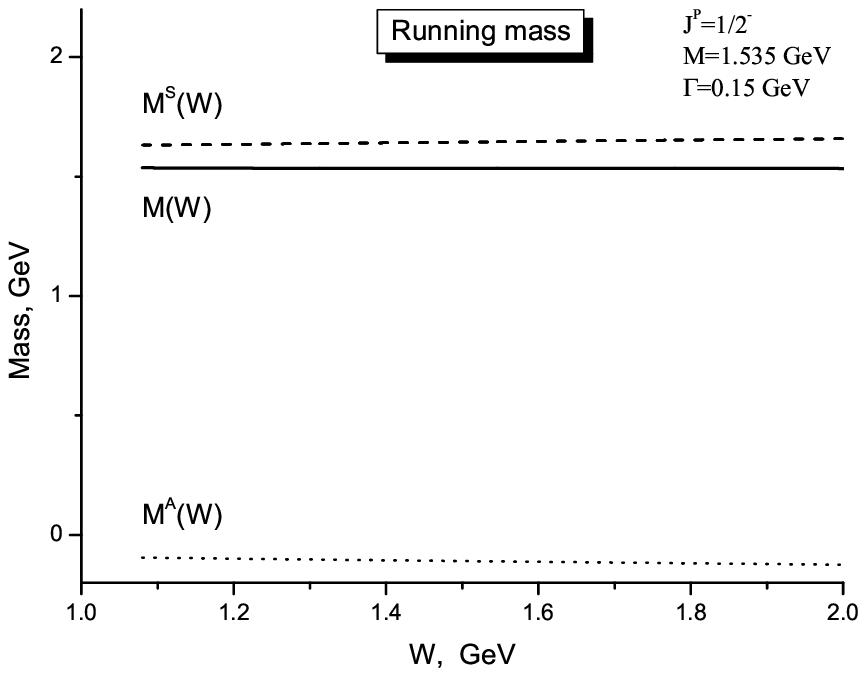}
  \includegraphics*[width=8cm]{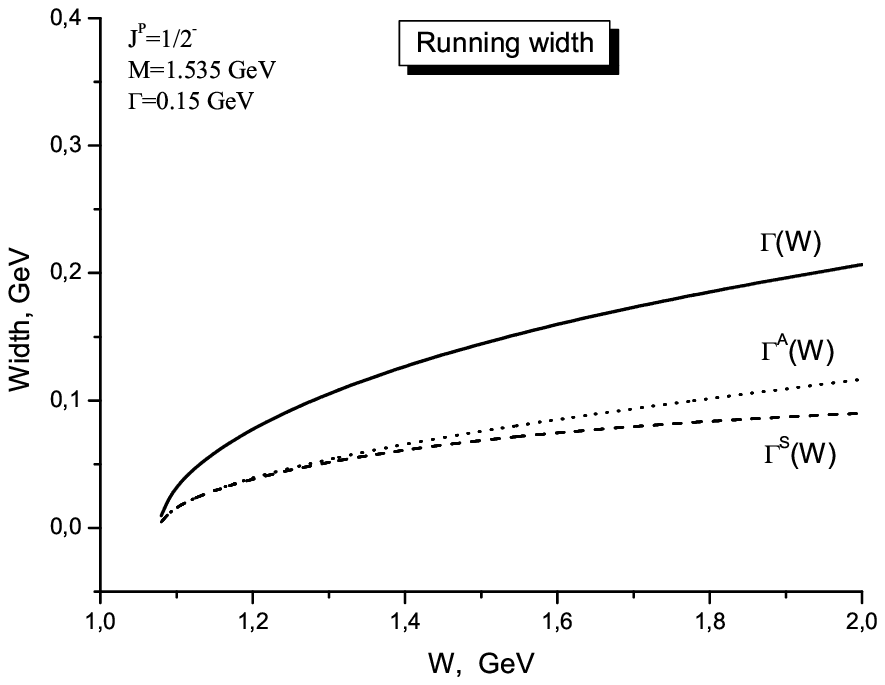}
  \caption{\label{fig2}Running mass and width and their components for resonance of
negative parity.}
\end{figure}

Figures \ref{fig1},\ref{fig2} demonstrate behaviour of the running
mass and width and its separation into symmetric and antisymmetric
parts. For illustration we considered production of $1/2^{\pm}$
resonance in $\pi N$ collisions with parameters resembling the
$N(1440)$ and $N(1535)$ baryons%
\footnote{In fact the $N(1440)$, $N(1535)$ do not look as the
isolated resonances (see  e.g. results of partial analysis
\cite{Arn04}).}%
. One can see that symmetric and antisymmetric parts of running
mass and width are of the same order and this is a typical
situation.

\subsection{Discussion}

We would like to discuss some limit cases in dressed fermion
propagator \eqref{BWF}. But first of all let us look at unitarity
condition and possible constrains from it. For definiteness, let
us consider $J=1/2$ resonance production in the $\pi N\to \pi N$
process and construct the partial waves. We use the standard $\pi
N$ kinematics \cite{Gaz} and \cite{Hoh,Cut,Arn04}.

\vspace*{1em}
\noindent\underline{\textbf{Resonance of positive parity}}
\vspace*{1em}

Effective lagrangian is of the form:
\begin{equation*}
  \Lagr_{int}=ig\big[\bar{N}^\prime(x)\gamma^5
  N(x)\cdot\varphi(x)+\bar{N}(x)\gamma^5 N^\prime(x)\cdot \varphi^\dag(x)\big]
\end{equation*}
Corresponding partial waves:
\begin{equation}
  \begin{split}
    f_{s,+}(W)&=\dfrac{-g^2\dfrac{(p^0+m_N)}{8\pi W}}{W-(M^S+M^A)+i(\Gamma^S+\Gamma^A)/2} =
               \dfrac{-(\Gamma^S+\Gamma^A)/p}{W-(M^S+M^A)+i(\Gamma^S+\Gamma^A)/2},\\
    f_{p,-}(W)&=\dfrac{g^2\dfrac{(p^0-m_N)}{8\pi W}}{-W-(M^S-M^A)+i(\Gamma^S-\Gamma^A)/2} =
               \dfrac{(\Gamma^A-\Gamma^S)/p}{-W-(M^S-M^A)+i(\Gamma^S-\Gamma^A)/2}.
  \end{split}
\end{equation}
Here $p^0$, $\vect{p}$ are the c.m.s. energy and momentum of
nucleon respectively.

\vspace*{1em}
\noindent\underline{\textbf{Resonance of negative parity}}
\vspace*{1em}

Effective lagrangian:
\begin{equation*}
  \Lagr_{int}= g \big[ \bar{N}^\prime(x) N(x)\cdot \varphi(x)+\bar{N}(x) N^\prime(x)\cdot \varphi^\dag(x) \big]
\end{equation*}
Partial waves:
\begin{equation}
  \begin{split}
    f_{s,+}(W)&=\dfrac{g^2\dfrac{(p^0+m_N)}{8\pi W}}{-W-(M^S-M^A)+i(\Gamma^S-\Gamma^A)/2} =
             \dfrac{(\Gamma^A-\Gamma^S)/\abs{\vect{p}}}{-W-(M^S-M^A)+i(\Gamma^S-\Gamma^A)/2},\\
    f_{p,-}(W)&=\dfrac{-g^2\dfrac{(p^0-m_N)}{8\pi W}}{W-(M^S+M^A)+i(\Gamma^S+\Gamma^A)/2} =
             \dfrac{-(\Gamma^A+\Gamma^S)/\abs{\vect{p}}}{W-(M^S+M^A)+i(\Gamma^S+\Gamma^A)/2}.
  \end{split}
\end{equation}
One can convince yourself that the partial amplitudes constructed
using dressed propagator satisfy the unitarity condition (last
expressions for partial waves demonstrate it explicitly)
\begin{equation*}
  \Im f = \abs{\vect{p}} \abs{f}^{2}
\end{equation*}
and it leads to
\begin{equation}
  \Gamma^A(W) > \Gamma^S(W) .
\end{equation}
So we can conclude that "quasi-bosonic" case $M^A=\Gamma^A=0$ in
the dressed propagator \eqref{BWF} is forbidden by unitarity
requirement.

The obtained field theory resonance contribution \eqref{BWF}
contains another interesting case, which may be called as
"anti-bosonic": $M^S=\Gamma^S=0$. In this case both partial waves
should coincide
\begin{equation}\label{ABos}
  f_{s,+}(W)=f_{p,-}(W)=-\dfrac{\Gamma^A/\abs{\vect{p}}}{W-M^A+i\Gamma^A/2}.
\end{equation}
This relation can not be true at all energies because these
partial waves have different orbital momentum and therefore should
have different threshold behaviour. But "anti-bosonic" case can be
realized as approximate equality in a resonance vicinity
\begin{equation*}
  f_{s,+}(W)\approx f_{p,-}(W)\quad \text{at}\quad W\sim M .
\end{equation*}
So we have degenerate resonances of different parity (it resembles
the restoration of chiral symmetry) but this physical situation is
realized in unusual manner: with using one spinor field $\psi(x)$.
In this case both components of a dressed propagator have
resonance nature
\begin{equation}\label{ABos1}
  G(p) \approx \Lambda^{+}\dfrac{1}{W-M+i\Gamma/2} - \Lambda^{-}\dfrac{1}{W-M+i\Gamma/2}\quad \text{at}\quad W\sim M .
\end{equation}

\section{Conclusion}

We investigated in detail the QFT improved Breit-Wigner formula for
fermions, obtained by means of the Dyson summation. The final
expression \eqref{BWF} for dressed propagator looks rather unexpected,
in particular, the resonance denominator differs from the well known
boson analog. This difference arises due to presence of antisymmetric in $W$
contributions in self-energy.

One can use the dressed fermion propagator \eqref{BWF} for hadron
phenomenology and in this case, after calculation of self-energy and
renormalization, we have two parameters: mass and width. But if we want
to obtain from \eqref{BWF} some simple parametrization for $W \sim M$
region, we need at least four parameters, as it can be seen from
the amplitude \eqref{intf}.

The essential technical ingredient of our consideration is the
using the off-shell projection operators. They simplify
essentially all calculations and clarify their physical meaning.
Recall that these projection operators were successfully used in
calculations of the $\Delta$-isobar propagator both in vacuum
\cite{KL,KL1} and media \cite{Kor}.

We thank V.M.Leviant for reading the manuscript and useful
remarks. This work was supported in part by Russian Foundation of
Basic Research grant No 05-02-17722.

\newpage

\end{document}